\documentstyle[12pt]{article}
\begin{document}
\def\bfl{\begin{flushleft}}
\def\efl{\end{flushleft}}
\def\bfr{\begin{flushright}}
\def\efr{\end{flushright}}
\def\beq{\begin{eqnarray}}
\def\eeq{\end{eqnarray}}
\def\nn{\nonumber\\}
\def\PRL{Phys. Rev. Lett.}
\def\PR{Phys. Rev.}
\def\RMP{Rev. Mod. Phys.}
\def\NP{Nucl. Phys.}
\def\PL{Phys. Lett.}
\def\REP{Phys. Rep.}
\def\CMP{Comm. Math. Phys.}
\def\JMP{J. Math. Phys.}
\def\CGG{Class. Quant. Grav.}
\def\MPL{Mod. Phys. Lett.}
\def\IJMP{Int. J. Mod. Phys.}
\def\AP{Ann. Phys.}
\begin{titlepage}
\rightline{CERN-TH/98-09}
\rightline{hep-th/9801064} \vspace{12pt}
\begin{center}\LARGE{\bf{\sf Effectiveness of One-Dimensional Gas Models
for Black Holes}}\\[24pt]
\Large{\sf A. Ghosh}\\
\normalsize{CERN\\
Theory Division\\
CH-1211 Geneva 23, Switzerland}\\[48pt]
{\Large Abstract}\\[12pt]
\end{center}
A one-dimensional gas model has been constructed and is shown to provide 
correct expressions for entropies for extremal and near-extremal BTZ black 
holes. Recently suggested boosting of black strings is also used to compute 
the entropy for the Schwarzschild black hole from this gas model.

\vspace{3cm}
\begin{flushleft}
CERN-TH/98-09\\
January, 1998
\end{flushleft}
\end{titlepage}
\section{Introduction}
In this note we attempt to propose a free one-dimensional bosonic gas  
model to describe the relevant degrees of freedom of a black hole, based on
some previous attempts, which addressed the cases of some specific black holes 
[1,2]. We show that it is possible to identify this way of counting states
with the counting procedure in a conformal field theory when mappings
are prescribed between the appropriate parameters in a conformal field
theory and the gas model. Note that no supersymmetry has been invoked in 
this model.

BTZ black holes in 2+1 dimensions provide a good example of attempting
to understand the black hole entropy by counting the microstates near its 
horizon [3]. Subsequently exact conformal field theories have been
constructed for this case [4]. Recently a class of geometries have also
been identified which share features with the BTZ case [5]. We
show here that the extremal and near-extremal BTZ black holes can be
adequately described by an appropriate one-dimensional gas of bosons by
computing its entropy from the gas model. In some sense, it reconfirms all
the previous calculations of the counting of states. The gas model cannot
predict the location of the states for the black hole. We also 
describe the importance of introducing a ground state advocated by many 
authors [4,5] in the context of constructing the conformal field
theory description for BTZ black holes. In the context of the gas model, this
will appear in a natural way and will provide a better starting point
for its implementation. We will make use of this concept to make an
improvement of the conventional gas model picture. The situation, however,
can hardly be improved for (3+1)-dimensional Reissner-Nordstrom black holes
or Schwarzschild black holes. We investigate this situation in some detail
in section 5. We also pointed out the essential objections against the
flat Minkowski space being treated as the possible `ground state' for most of 
the asymptotically flat black holes in 3+1 dimensions. 

There have been many attempts [6] to provide a counting of states for the
Schwarzschild black holes in various dimensions, including recently some 
from M-theory. The key idea is to map the Schwarzschild black hole to extremal
or near-extremal configurations from which the counting might be possible.
One concrete realization of this mapping was provided by [7], where it is shown
that black branes/strings give rise to Schwarzschild or charged dilatonic 
black holes when reduced to lower dimensions before or after boosting along 
some uncompactified direction. This mapping has been used to argue even the 
matching of entropies for these two distinct classes of black holes [8]. Some
counting has also been performed for seven-dimensional Schwarzschild
black holes, by computing the entropy of a near-extremal three-brane obtained
by this boosting and dualizations [9]. We provide such a boosting on
a five-dimensional black string to map to a charged dilatonic black hole.
Then a gas model computation of the entropy has been performed, which
eventually predicts, by construction, the entropy of a Schwarzschild black
hole. In this way in 3+1 dimensions we found an exact agreement between 
the gas model and the Schwarzschild black hole entropy.

We first describe the model and its relation to conformal field theories
in sections 2 and 3. In subsequent sections we apply it to various black 
holes in various dimensions. Starting with 
the (2+1)-dimensional BTZ in section 4, we put forward an improved version of 
the model in section 5. Schwarzschild black holes in 3+1 dimensions are
discussed in section 6.
 
\section{The gas model}
In this section I shall describe the model.
Let us take a one-dimensional gas consisting of $N$ massless
bosons in thermal equilibrium at temperature $T$. The space is a box of
length $L$ with periodic boundary conditions at the ends. We have used 
physical units throughout and the Newton constant, according to our 
conveniences, is taken to have various values, which will be specified later on. The spectrum of the model consists of an infinite number of oscillators with 
energy eigenvalues 
\beq
E_n=n\omega,\qquad n\in{\bf Z},\omega={2\pi\over L}.\eeq
The free energy of this system is given by the partition function
\beq
\beta F&=&\sum_n\log[1-\exp(-\beta E_n)]\nn
&\simeq&\int_0^{\infty}dn\log[1-\exp(-\beta\hbar n\omega)]
=-{\pi^2\over 6\beta\omega},\eeq
i.e.
\beq
F=-{\pi T^2L\over 12}.\eeq
All states being massless, there will be in general left-moving and 
right-moving 
modes for which we put indices $L$ and $R$. The expressions for free energy, 
entropy and total energy are the following
\beq
F_{L,R}&=&-{\pi\over 12}LN_{L,R}T^2_{L,R}\nn
S_{L,R}&=&-\partial F_{L,R}/\partial T_{L,R}={\pi\over 6}LN_{L,R}T_{L,R}\nn
E_{L,R}&=&F_{L,R}+T_{L,R}S_{L,R}={\pi\over 12}LN_{L,R}T_{L,R}^2\eeq 
with left (right) oscillators are in equilibrium at temperature $T_{L(R)}$.
Combining left and right sectors, the total energy and momentum of the system 
will be given by
\beq
E&=&E_L+E_R+E_0\nn
P&=&E_L-E_R,\eeq
where $E_0$ is the thermal ground state energy correction, which we have 
left out in (2.1).
Throughout the calculation we shall keep the net momentum $P$ and ground 
state energy $E_0$ fixed for the system. The reason for doing this will be
clarified when we come back to this point. The immediate 
consequences are
\beq
dP=dE_L-dE_R=0,~dE_0=0\Rightarrow dE=2dE_L=2dE_R.\eeq
So the net temperature $T$ is given by the formula
\beq
{1\over T}={\partial S\over\partial E}={\partial S_L\over 2\partial E_L}+
{\partial S_R\over 2\partial E_R}={1\over 2}\left({1\over T_L}+{1\over T_R}
\right).\eeq
These are all the formulas that we need for our present purpose. We shall 
apply this model to describe excitations of a generic black hole over a 
corresponding `ground state'. The meaning and importance of choosing this 
`ground state' will be clarified later. Our strategy will be the following:
we shall identify the temperature and the energy of the gas with the
corresponding objects for black holes (by energy of the black hole we mean
its ADM energy). Then we shall show that the entropy of the gas model predicts
exactly the entropy of the black hole.  At this point we shall see how the 
entropy of the gas model compares with the exact conformal field theory(CFT) 
counting.
 
\section{Relation to CFT}
Here we discuss how the model described in the previous section
relates to CFT. The basic motivation would be to try to identify this model 
with some conformal field theory whose left and right sectors are to be 
realized as the left and right oscillators of this model. First we define
the central charge from the gas model to be $c=N_L=N_R$. Now let us take a 
primary state in the CFT at level ($h_L,h_R$) and associate this with 
the quantum mechanical state of a black hole:
\beq
|h_L,h_R\rangle\equiv|{\rm black~hole}\rangle\eeq
Clearly, the underlying assumption is that the quantum theory of gravity, in a
restricted sense, is an exact CFT. This does not sound unnatural at present
as many examples have already been constructed and `verified' [3,10].
In CFT the degeneracy of this state (for $h_{L,R}\gg 1$) is known to be
\beq
d(h_L,h_R,c)\simeq\exp\left[2\pi\sqrt{ch_L\over 6}+2\pi\sqrt{ch_R\over 6}
\right].\eeq
So the Boltzmann entropy of this state is given by
\beq
S_B=\log d(h_L,h_R,c)\simeq 2\pi\sqrt{ch_L\over 6}+2\pi\sqrt{ch_R\over 6}.\eeq
Now for the gas model if we define the levels $h_L,h_R$ by
\beq
P=2\pi{h_L-h_R\over L}\eeq
then immediately we get
\beq
S_B={\pi L\over 6}[N_LT_L+N_RT_R]=S,\eeq 
where we have used $c=N_L=N_R$. The implication of the above
result is that whenever the gas model entropies would give rise to the correct
expressions for the black hole entropies then there always exists a natural,
underlying CFT that can also be used to describe the black hole states.
In this note we shall give evidence of this implication
only by computing black hole entropies. Of course, more evidence should be
collected by computing other objects, such as emission and absorption cross
sections of black holes. Note, however, that this identification of
$S_B=S_{gas}$ is independent of whether the gas model is applied to 
black holes or not.

\section{Entropy of BTZ black holes}
Let us consider the case of the three-dimensional BTZ black hole for
which the entropy has already been calculated, using exact conformal field
theories [3,4]. Here we are doing this exercise just to provide 
more evidence
to the parallelism of the gas model and CFTs. It is important to notice 
that this does not require the existence of supersymmetry in support of the 
correctness of the counting or validity of the formulation. We shall take up
the case of the near-extremal BTZ black hole. The reason is that, for
exact non-extremal BTZ black hole, the identifications of temperature and
energy with those of the gas model do not give sufficient information
about the model, so that we can compute the entropy. However, for the black
hole near to its extremality some detailed structure of the gas model can
be used to completely specify all the relevant parameters for computing its
entropy from just the temperature and energy equations. In this section
we shall first use the extremal black hole to fix some gas model parameters.
In the next section we shall improve on this procedure to predict the
near-extremal entropy from more elementary considerations.

Gravity in $2+1$ dimensions with a cosmological constant $\Lambda=-1/l^2$,
furnishes rotating black hole solutions (BTZ black hole) [11]
\beq
ds^2=-f^2dt^2+f^{-2}dr^2+r^2\left(d\phi-{J\over 2r^2}dt\right)^2,
\eeq
where $f^2=-M+r^2/l^2+J^2/4r^2$. The two horizons, the temperature and the 
entropy of this black hole are given by
\beq
r_{\pm}^2&=&{1\over 2}Ml^2\left[1\pm\left(1-{J^2\over M^2l^2}\right)^{1/2}
\right]\nn
T_H&=&{r_+^2-r_-^2\over 2\pi r_+l^2}\nn
S_H&=&4\pi r_+,
\eeq
where we have taken the Newton constant to be $G_N=1/8$. The extremal limit 
of this black hole amounts to identifying the two horizons $r_+=r_-$ or 
$J=Ml$.

Now let us use the extremal black hole parameters to restrict the gas model
to some extent. We shall equate energy, temperature and entropy to 
determine some parameters of the gas model. For the extremal solution
\beq
T_H=0, E_H=|J|/l, S_H=4\pi\sqrt{Jl/2}.
\eeq
First, equating the temperatures one finds that there are two possibilities:
a) both left and right temperatures are zero and b) one of the two is zero.
But the possibility (a) can be discarded by considering the entropy. 
Let us take $T_R=0\Rightarrow E_R=0$. So
\beq
{J\over l}&=&E_0+{\pi LN_L\over 12}T^{(0)2}_L\nn
4\pi\sqrt{Jl/2}&=&{\pi LN_L\over 6}T^{(0)}_L,\eeq
where the superscript $(0)$ stands for objects in the extremal case. 
Equations (4.16) give rise to
\beq
T^{(0)}_L&=&{24\over N_LL}\sqrt{Jl\over 2}\nn
E_0&=&{J\over l}-{24\pi Jl\over N_LL}\nn
P&=&{J\over l}-E_0.\eeq
Now let us feed it to the near-extremal case for which $M=J/l+\epsilon,
\epsilon\ll J/l$. The two horizons, the Hawking temperature and the entropy 
are given by
\beq
r_{\pm}&\simeq&\sqrt{Jl\over 2}\left(1\pm\sqrt{\epsilon l\over 2J}+{\epsilon 
l\over 2J}\right)+O(\epsilon^{3/2})\nn
T_H&\simeq&{1\over l\pi}\left(\sqrt\epsilon-\epsilon\sqrt{l\over 2J}\right)
+O(\epsilon^{3/2})\nn
S_H&\simeq&4\pi\sqrt{Jl\over 2}+2\pi l\sqrt\epsilon+\pi l\epsilon\sqrt{
l\over 2J}+O(\epsilon^{3/2}).
\eeq
In the gas model we shall keep $P$ and $E_0$ fixed (their values
correspond to the BTZ class in the sense that they are fixed by the extremal 
solution alone; similar observations have been made recently for the case
of dyonic black holes too [12]). This gives rise to the following 
expressions for $E_L,E_R$
\beq
E_L+E_0&=&{J\over l}+{\epsilon\over 2}\nn
E_R&=&{\epsilon\over 2},\eeq
so that $E=E_L+E_R+E_0=J/l+\epsilon=M$ and $P=E_L-E_R=J/l-E_0$.
From the above identifications we find $T_L$ and $T_R$ to be (using $E_0$)
\beq
T_L&\simeq&{24\over LN_L}\sqrt{Jl\over 2}+{\epsilon\over 4\pi\sqrt{2Jl}}
+O(\epsilon^{3/2})\nn
T_R&=&\left({6\epsilon\over \pi LN_R}\right)^{1/2}.\eeq
So the net temperature of the gas is given by
\beq
T={2T_LT_R\over T_L+T_R}\simeq 2\left({6\epsilon\over \pi LN_R}
\right)^{1/2}-{\epsilon
N_L\over l\pi N_R}\sqrt{l\over 2J}+O(\epsilon^{3/2}).\eeq
Comparing this temperature with $T_H$ the $\sqrt\epsilon$-terms give a value 
for $LN_R$ and the $\epsilon$-terms give a value for $LN_L$:
\beq
LN_L=LN_R=24\pi l^2.\eeq
If we put all these values in (2.15) and (2.17) then the entropy becomes
\beq
S={\pi L\over 6}(N_LT_L+N_RT_R)=S_H
\eeq
up to order $\epsilon$. At this point let me also mention the relation
of these calculations with [3]. It is natural to take $L=l$, the scale
set by the cosmological constant. Then the central charge of the corresponding
CFT is given by $c=24\pi l$, in agreement with [3]. If we identify the
levels from their definitions (3.11) and put them in (3.10), then using the
above central charge the entropy formula in (4.18) is easily reproduced.

\section{A detour}
In this section we make an improvement of the naive implementation
of the gas model made in the previous section. As an example, consider again
the case of the BTZ black hole. Of course this detour follows the basic 
procedure adopted in the previous section, but we will see that this can 
reproduce the results obtained in section 4 more independently. The 
philosophy is the following: we shall first
identify a suitable `ground state' for the black hole and eventually 
build up our black hole over that ground state in a systematic way. 
The corresponding gas model description will follow automatically 
from this construction. 

For BTZ let us start from the state described by the configuration
$M=J=0$
\beq
ds^2=-{r^2\over l^2}dt^2+{l^2\over r^2}dr^2+r^2d\phi^2.\eeq
The relevance of this `ground state' is also described in [4,5],
where it is identified as the ground state in the Ramond
sector of a super-CFT whereas the adS-space-time is identified as the
ground state in the Neveu-Schwarz sector. The CFT relevant to the BTZ black 
hole is equivalent to the quantum theory of gravity built on the adS vacuum. 
We will not invoke any supersymmetry here and show, in the context of the 
gas model, that excitations over this ground state alone are adequate for 
describing the near-extremal BTZ configuration. For this state we have [4], 
$E_H=T_H=S_H=0$. From the particle model point of view, both $T_L,T_R$
should be vanishing as the entropy is zero in this case, from which
we conclude that the ground state energy $E_0=0$. Also the important
thing to notice is that for this configuration the left and right sectors
of the particle model behave democratically. Now over this
state we want to build up the extremal BTZ configuration.  
One way would be to throw an arbitrary amount of matter such that 
$M=J/L$ always holds. We know from our analysis in the previous section  
that for this configuration the particle model stops behaving democratically
in its left and right sectors. The basic reason is that one cannot put
both $T_L,T_R$ to zero, because the extremal black hole has a finite entropy.
The right sector still remains at its ground state and the left sector 
picks up excitations (this was our convention in the earlier section). Then
at once we can identify the energy and the entropy to be
\beq
S=2E_L/T_L,\qquad{\rm where}~E_L=J/l.\eeq
We do not learn anything more than this as $T_L$ remains an
arbitrary parameter. On the contrary, we will now see how all the
information can be extracted if we relax the extremality condition 
a little bit. This is more natural from the point of view of how
the extremal configurations are actually obtained: first throw an arbitrary
amount of matter such that we do not care about the extremality
holding exactly, but it is sufficient to restrict to a near-extremal 
configuration and let $\epsilon\to 0$ at the end of the process. So we 
let our temperatures be
\beq
T_R&=&a_1\sqrt\epsilon+a_2\epsilon+...\nn
T_L&=&T_0+b_1\sqrt\epsilon+b_2\epsilon+...\eeq
We will fix these unknown constant coefficients $T_0,a_i,b_i$'s by comparing
gas model temperature and energy with the corresponding objects for the
black hole. The temperature identification immediately gives
\beq
a_1={1\over 2\pi L},\quad a_2=0,\quad T_0={1\over\pi l^2}\sqrt{Jl\over 2};\eeq
$a_2=0$ does not follow automatically from this, but it turns out 
to be zero by keeping $E_0=0$ fixed up to order $\epsilon$.
We still use the basic decomposition equations (4.19) to fix the
other constants (when $P,E_0$ are kept fixed):
\beq
b_1=0,\quad b_2={1\over 4\pi\sqrt{2Jl}},\quad LN_L=LN_R=24\pi l^2.\eeq
So we have reproduced all the information about the
gas model required to compute the entropy. Recall that here we
did not make use of the energy, the temperature or the entropy of the
extremal configuration explicitly. This is why this approach is more 
powerful. 

When applied to other cases the situation is not so
robust. One needs to fix a funny ground state energy $E_0$ by
making use of the extremal configuration explicitly. Then we are back
to the level of the naive gas model. The situation can be
demonstrated for Reissner-Nordstrom black holes in four dimensions [1,2].
If one performs similar steps for this black hole then, before making
use of the energy decomposition equations to fix the other constants, we can 
go back and compute the entropy for the extremal configuration since, for 
doing this, it is adequate to 
know the value of $T_0$. Then one sees that we need a finite value for
$E_0$ to match everything consistently. With standard normalizations $T_0=
1/8\pi Q$ and $E_0=15Q/16$, where $Q$ is the Reissner-Nordstrom charge.
Making use of this funny shift in ground state energy, everything fits
appropriately for extremal and near-extremal configurations. But for
choosing a `ground state', much like the $M=J=0$ state
or the adS-state for the BTZ black hole, it is natural to pick up the
state $M=Q=0$ for Reissner-Nordstrom black hole, which is the flat Minkowski
space-time. This state should have zero temperature and zero entropy. 
In the gas model, $T=S=0$ implies that the 
ground state energy $E_0=0$. Now if one performs similar steps as for
the BTZ case then one comes up with wrong prefactors in the 
entropy. Of course, the qualitative dependence on area always comes out, 
but we are not satisfied with this for a reasonable model. This certainly
destroys the predictability of the improved approach advocated above. 
To our feeling, this is due to the wrong choice of the 
`ground state', which for Reissner-Nordstrom is taken to be the Minkowski 
space. A correct choice would certainly furnish an improved gas model
description of the Reissner-Nordstrom black hole. For the case of the
Schwarzschild black hole, however, such an extremal counterpart is missing 
and, hence, even a naive particle model computation of entropy is not 
possible. If, however, one attempts to build up the Schwarzschild black hole
on the state $M=0,S=0,T=\infty$, then again the prefactors do not come
out correctly. A detailed comparison between the gas model and the 
Schwarzschild black hole parameters reveals the fact that in this case too
a funny ground state energy is needed.  

\section{Entropy of Schwarzschild black hole}
In this section we consider the case of the Schwarzschild black hole. 
We will show how at least a naive gas model description can
be provided in this case. The basic idea that we are going to use is
the following:

One starts from a five-dimensional black string [7]:
\beq
ds_5^2=-\left(1-{2m\over r}\right)dt^2+\left(1-{2m\over r}\right)^{-1}dr^2+
r^2d\Omega_2+dy^2,\eeq
where $y$ represents a flat direction. When compactified in the flat direction
using the standard Kaluza-Klein reduction, this provides a Schwarzschild black 
hole in four dimensions with mass $m$ and entropy $4\pi m^2$, and we have set 
the four-dimensional Newton constant to unity. But if we first perform a
Lorentz boost in the uncompactified $t-y$-plane and subsequently compactify 
the boosted $y$, then we get a charged dilatonic black hole in four dimensions.
If the compactification radii in these two cases are related by $R_1=R_2\cosh
\alpha$, where $1,2$ refers to the Schwarzschild and charged black hole cases, 
respectively, and $\alpha$ is the boosting parameter, then the entropy of the 
non-extremal charged dilatonic black hole is identical to that of the 
Schwarzschild black hole [8]. Notice that $R_2$ is precisely the Lorentz
contracted radius of $R_1$ under boosting. Certainly a boost along a compact 
internal dimension is not a symmetry of the solution but the entropy should be 
the same, since one expects that the number of states should not change under 
boosting [13,8]. Then it is reasonable to count the number of states 
for one classical configuration and predict the entropy of the other when two 
configurations are related to each other by boosting [9]. This is precisely
what has been done in the following part of this section. A counting of 
states is performed for the fully non-extremal (we will not subject ourselves 
to a near-extremal limit in this case) charged dilatonic black hole to 
find out the entropy of the Schwarzschild black hole. One should note that 
for every compactification that gives rise to the Schwarzschild black hole 
there exists a corresponding compactification radius that gives rise to a 
charged dilatonic black hole and, hence, this identification makes sense.
We will come back to the point of black string - black hole transition at 
the end. The problem of counting for these charged dilatonic black holes is 
that they usually have a singular BPS limit, in the sense that there is a 
curvature singularity at the horizon and the area of the horizon goes to 
zero in the extremal limit [14]. However, the thermodynamics of these 
black holes have been carried out in detail in the past [15,16]. 
Let us now provide a gas model counting of states for such black holes. We
will use the convention used in [17,16].

The generic black hole is ($G=1$)
\beq
ds^2=-{r^2-2mr\over\sqrt\Delta}dt^2+{\sqrt\Delta\over r^2-2mr}dr^2+
\sqrt\Delta d\Omega_2,\eeq
where $\Delta=r^2[r^2+2mr(\cosh^2\alpha-1)]$ and the dilaton is given by
\beq
\phi=\ln(r^2/\sqrt\Delta).\eeq
The ADM mass and the charge are given by
\beq
M&=&{1\over 2}m(1+\cosh^2\alpha)\nn
Q&=&{1\over\sqrt 2}m\sinh\alpha\cosh\alpha.\eeq
The Hawking temperature and entropy are given by (the horizon is at $r=2m$):
\beq
T_H^{-1}=8\pi m\cosh\alpha,~S_H=4\pi m^2\cosh\alpha.\eeq
The BPS limit of this black hole amounts to taking $m\to 0,~\alpha\to\infty$,
keeping $m\cosh^2\alpha=m_0$ finite. In the extremal limit, $S_H=0,~T_H=
\infty,~M=m_0/2,~Q=m_0/\sqrt 2$, which satisfies the BPS condition $M^2=Q^2
/2$.

Now comparing the gas model temperature, energy and entropy with those of
the extremal black hole, we fix the following parameters:
\beq
T=\infty,\quad N_L=N_R=0,\quad E_0={m_0\over 2}\eeq
Note that both the left and right temperatures need to be infinity 
in this case. Also the left and right sectors are behaving 
democratically in this case. There is a similarity with the $M\to 0$ limit 
of the Schwarzschild black hole, that the `ground states' in both cases
correspond to $S=0,~T=\infty$.
The important difference is that there is a finite ground state energy
in this case, which we have already mentioned to be significant for describing
the Schwarzschild black hole. Also note that for this extremal black hole 
$P=0$.
 
The non-extremal deformation is tuned by turning the parameter $m$ to
a finite value. Let us subscribe to a specific case for which a gas model
description can be obtained easily. So we choose a curve in the ($m,\alpha$)
-plane, given by the equation $m\cosh^2\alpha=m_0-m/2$. This amounts to
specifying a relation between the ADM mass and the charge. Note that this is
a particular non-extremal deformation over the BPS solution (6.34) which is
tuned by the parameter $m$. In the corresponding Schwarzschild case $m$
turns out to be the ADM mass. The ADM mass, temperature and entropy for this
non-extremal black hole, when expressed in terms of the parameter $m_0$, 
are given by
\beq
M&=&{m_0\over 2}+{m\over 4}\nn
T_H^{-1}&=&8\pi\sqrt{mm_0-m^2/2}\nn
S_H&=&4\pi m\sqrt{mm_0-m^2/2}.\eeq
Also the charge becomes 
\beq
Q={m_0\over\sqrt 2}\left(1-{m\over 2m_0}\right)^{1/2}\left(1-{3m\over
2m_0}\right)^{1/2}\eeq
and hence the BPS condition $M^2=Q^2/2$ no longer holds.
If we use the left-right democracy $N_L=N_R=N,~T_L=T_R=T$ then it is possible
to study the full non-extremal case rather than the near-extremal one.
The number of parameters become half in this case, and the temperature
and energy equations are sufficient to fix the rest of the parameters of the
gas model. Equating the temperatures and ADM mass with the energy of the 
gas we get
\beq
T_H=T,\qquad LN=96\pi m^2(m_0-m/2).\eeq
So the gas-model entropy is given by $S=4\pi m\sqrt{mm_0-m^2/2}$, in perfect
agreement with the non-extremal entropy of the charged dilatonic black
hole, and hence, when boosted, with that of the Schwarzschild black hole. It
is also instructive to calculate the central charge of the underlying
CFT, as we did for the case of BTZ black hole. We choose the scale $L$, here,
to be $m_0-m/2$ characteristic to the curve we have chosen. Thus,
\beq
c=96\pi m^2.\eeq
If one identifies the level from the definition (3.11) then (3.10) gives 
the desired expression for entropy. At this point one should note that
with this central charge one can calculate the entropy for the Schwarzschild
the black hole directly. Again we take a left-right symmetric gas with
$T=1/8\pi m$ in units where $G=1$. Also a direct extremal limit being
absent the natural scale is set by the black hole mass, $L=m$. Then 
counting formula (3.12) gives rise to the entropy: $S=(\pi cL/3)T=4\pi m^2$,
in agreement with that of the Schwarzschild black hole.

So we conclude that there exists a CFT description for the non-extremal
charged dilatonic black holes, which have singular extremal limit. 
With the identification of the compactification radii these calculations 
indicate that there exists a CFT for the Schwarzschild black hole too
for which we have calculated the central charge $c=96\pi m^2$. 
At this point let me mention some subtleties involved in the identification
of black hole entropies with two compactification radii.
In the extremal limit the boosting is infinite and hence $R_1\gg R_2$.
In the region where $R_1$ is larger than the Schwarzschild radius the black 
hole should be viewed as a black string and what we are computing is basically
the entropy of the black string. So the extremal or near-extremal charged 
dilatonic configurations basically describe the back string entropy rather 
than the hole entropy. To really probe into the black hole region one should
compute the entropy for configurations obtained by finite boostings.
This forces us to consider the charged dilatonic black holes
far from extremality. For a non-extremal charged dilatonic black 
hole, finite $\alpha$ and small $R_2$ imply small $R_1$ ($R_1$ smaller than
the Schwarzschild radius), which describes the black hole region 
appropriately.  

\begin{center}
{\large\bf Acknowledgements}
\end{center}
The author would like to acknowledge discussions with Parthasarathi Mitra and
Jnan Maharana. This research was supported in part by the World Laboratory. 

\vspace{1cm}
\noindent Note: When this work was finished we saw a paper in the 
archive [18] also addressing the entropy of Schwarzschild black holes in 
four dimensions.

\end{document}